\documentclass[journal, twoside, final]{IEEEtran}

\usepackage{cite}
\usepackage{graphicx}
\usepackage{amssymb, amsmath, amsthm}
\usepackage{amsfonts}
\usepackage{mathrsfs}
\usepackage{color}
\usepackage{url}
\usepackage{flushend}
\usepackage{setspace}
\usepackage{placeins}
\usepackage{booktabs, multicol, multirow}
\usepackage{acronym}
\usepackage{bm}
\usepackage{placeins}
\usepackage{extarrows}
\usepackage{color}
\usepackage{makecell}
\usepackage[caption=false, font=footnotesize]{subfig}

\usepackage{pict2e}
\usepackage{algorithm, algorithmic}

\begin{document}

\title{Bayesian Compressive Channel Estimation for Hybrid Full-Dimensional MIMO Communications}

\author{Hongqing~Huang, Peiran~Wu,~\IEEEmembership{Member,~IEEE,} Minghua~Xia,~\IEEEmembership{Senior~Member,~IEEE}
	
	\thanks{The authors are with the School of Electronics and Information Technology, Sun Yat-sen University, Guangzhou 510006, China (e-mail: huanghq29@mail2.sysu.edu.cn, \{wupr3, xiamingh\}@mail.sysu.edu.cn).}
}

\maketitle

\begin{abstract}
Efficient channel estimation is challenging in full-dimensional multiple-input multiple-output communication systems, particularly in those with hybrid digital-analog architectures. Under a compressive sensing framework, this letter first designs a uniform dictionary based on a spherical Fibonacci grid to represent channels in a sparse domain, yielding smaller angular errors in three-dimensional beamspace than traditional dictionaries. Then, a Bayesian inference-aided greedy pursuit algorithm is developed to estimate channels in the frequency domain. Finally, simulation results demonstrate that both the designed dictionary and the proposed Bayesian channel estimation outperform the benchmark schemes and attain a lower normalized mean squared error of channel estimation.
\end{abstract}


\section{Introduction}
\label{Section-I}  

Millimeter-wave (mmWave) communications, especially the mmWave full-dimensional multiple-input multiple-output (FD-MIMO) communications, have become a dominant technique of the forthcoming 6G wireless systems due to their extremely wide bandwidth and high spatial degrees of freedom~\cite{Overview2016Heath}. To attain a tradeoff between hardware complexity and achievable data rate, implementing FD-MIMO in practice entails hybrid digital-analog architecture~\cite{swomp2018jp}. This architecture allows fewer radio frequency (RF) chains than antenna elements, reducing hardware complexity and cost. However, it poses a challenge to estimate channels from a limited number of pilots. Therefore, efficient channel estimation plays a pivotal role in implementing mmWave FD-MIMO communication systems.

As mmWave channels are typically sparse in the angular domain, many strategies based on compressive sensing (CS) and sparse recovery have been developed to estimate hybrid FD-MIMO channels by exploring the angles of arrival and departure. In particular, the most popular CS-based strategies include iterative algorithms~\cite{9298895} and greedy algorithms, e.g., orthogonal matching pursuit (OMP)~\cite{ompce2016Lee} and simultaneous weighted orthogonal matching pursuit (SWOMP)~\cite{swomp2018jp, SWOMP2018Javier}. Compared with the OMP, the SWOMP exploits multiple measurements to improve the estimation accuracy, yielding a lower convergence speed. Regarding the CS-based methods, redundant dictionaries consisting of array steering vectors with finely quantized angle grids are widely used~\cite{ompce2016Lee}. 

In companion with the applications of CS, sparse Bayesian learning (SBL) is applied recently for channel estimation~\cite{2017BL,2018BL,2021BL}. For the Bayesian estimators, one initializes a parameterized prior probability of the parameters to be estimated. Then, the expectation-maximization technique can be employed to approximate the optimal estimator of its hyperparameters. However, the SBL-based estimators have higher computational complexity, albeit more accurate, than the CS-based greedy estimators. Thus, it is natural to ask how the Bayesian inference is exploited in the CS-based greedy algorithms to improve accuracy with economic complexity.

Motivated by the spherical Fibonacci grid (SFG) \cite{Fibonacci2009Gonz} that provides uniform grids on a sphere and the Bayesian tools~\cite{booksparse}, this letter develops a channel estimation technique for hybrid FD-MIMO systems. Specifically, we first design a new dictionary based on the SFG in the CS framework, which provides minor angular errors than traditional dictionaries~\cite{offgrid2020Anjinappa}. Then, a Bayesian inference-aided greedy pursuit algorithm is developed for accurate channel estimation by integrating the Bayesian inference into the general CS framework. Finally, Monte-Carlo simulation results demonstrate the superiority of the developed channel estimation method.


\subsubsection*{Notation}
Vectors and matrices are denoted by lower- and upper-case letters in boldface, respectively. Vectors are by default in column, and the $i$-th entry of a vector is specified by the subscript $ [\cdot]_{i} $. The operators $\left(\cdot\right)^{-1}$, $\left(\cdot\right)^{T}$, and $\left(\cdot\right)^{H}$ indicate the inverse, transpose, and conjugate transpose, respectively. Also, $\det(\cdot)$ and ${\rm tr}(\cdot)$ denote the determinant and trace, respectively. The operators $\otimes$ and $\left\|\cdot\right\|_{\mathrm{F}}$ denote the Kronecker product and the Frobenius norm, respectively. The set of indices of non-zero entries in a sparse vector is called a support.

\section{System and Channel Models}
\label{Section-II}

\begin{figure}[t!]
	\centering
	\includegraphics[width=0.85\linewidth]{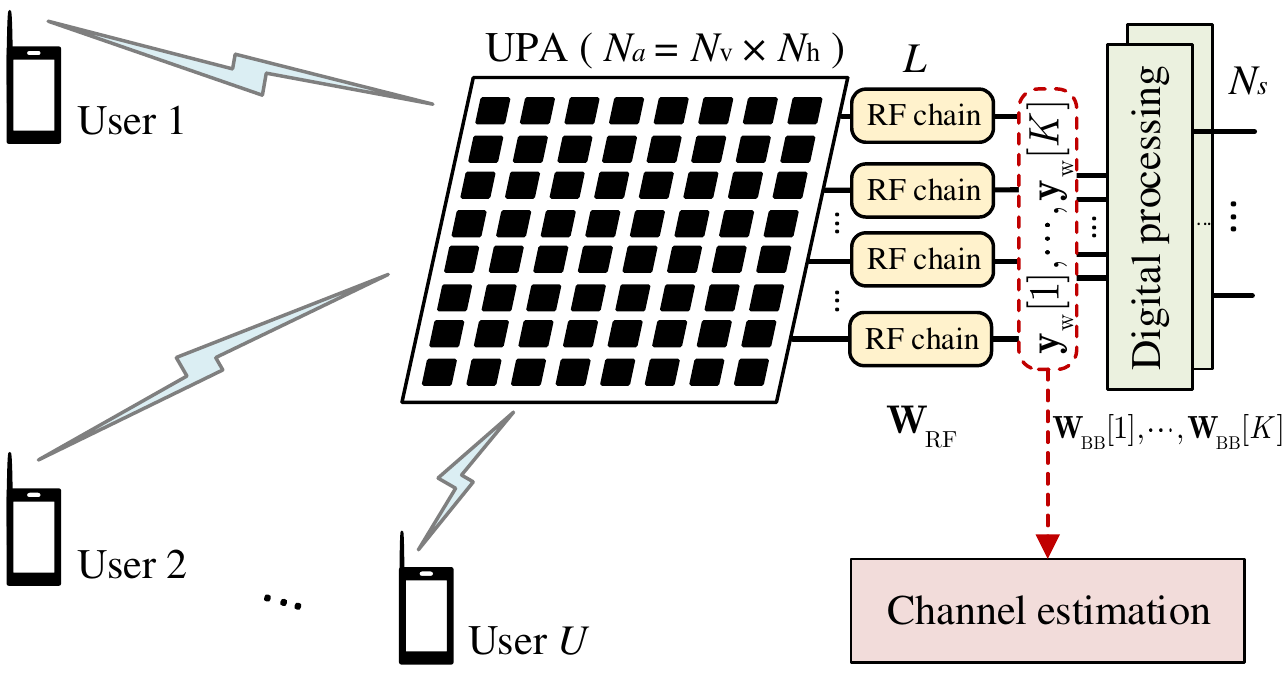} 
	\caption{A FD-MIMO communication system with a hybrid architecture.}
	\vspace{-8pt}
	\label{fig:1}
\end{figure}

Figure~\ref{fig:1} depicts a hybrid FD-MIMO system with a uniform planar array (UPA), $ K $ subcarriers, and $ N_{s} $ data streams. There are $U$ single-antenna users and a base station (BS) equipped with $L$ RF chains and a UPA composed of $N_{a} = N_{\rm v} \times N_{\rm h}$ antennas, with $N_{\rm v}$ and $N_{\rm h}$ being the number of antenna rows and columns, respectively. Regarding the $L$ RF chains, the analog precoder/combiner $\bm{W}_{\rm RF}\in\mathbb{C}^{N_{a} \times L}$, with unit modulus $|\left[\bm{W}_{\rm RF}\right]_{i, j}|=1$, for all $1\leq i\leq N_{a}$,~$1\leq j\leq L$,~are implemented by phase shifters with a fully connected architecture \cite{Phase2016Rial}. For the digital processing, the precoders/combiners are frequency-selective, i.e., $\{\bm{W}_{\rm BB}\left[k\right]\in\mathbb{C}^{{L}\times N_{s}}\}_{k=1}^{K}$ for $K$ subcarriers are distinct. In practice, we have $N_{s} \leq L < N_{a}$ in a hybrid FD-MIMO architecture~\cite{Overview2016Heath}. 

When the FD-MIMO system shown in Fig.~\ref{fig:1} operates in mmWave frequencies, the corresponding sparse channels can be described by using a geometry-based stochastic model~\cite{GBSM2018}. In particular, given the length $L_{\rm d}$ of delay taps in the time domain, the channel pertaining to the $l$-th delay tap, $l=1,\cdots,L_{\rm d}$, can be expressed as~\cite{SWOMP2018Javier}
\begin{equation}
	\bm{h}\left(l\right)=\sum_{n=1}^{N_{\rm p}}g_{n}f_{\rm rc}\left(lT_{\rm s}-\tau_{n}\right)\bm{a}\left(\theta_{n}, \phi_{n} \right), \label{eq:1}
\end{equation}
where $N_{\rm p}$ denotes the number of spatial scattering paths, $g_{n}$ is the complex-valued channel gain of the $n$-th path, $f_{\rm rc}\left(t\right)$ is a raised-cosine band-limited filter sampling at time $t$, $T_{\rm s}$ refers to the sampling period, $\tau_{n}$ is the time delay of the $n$-th path, and $\bm{a}\left(\theta_{n},\phi_{n}\right)\in\mathbb{C}^{N_{a}\times1}$ is the array steering vector corresponding to the $n$-th path, in which $\theta_{n}$ and $\phi_{n}$ indicates the zenith and azimuth angles of the $n$-th arrivial path, respectively.

Given the horizontal spacing $d_{\rm h}$ and vertical spacing $d_{\rm v}$ of any two antenna elements of the UPA under study, the normalized array steering vector with respect to a zenith $\theta \in \left[0,\pi\right]$ and an azimuth $\phi \in \left[-\pi,\pi\right)$ is determined by
\begin{equation}
	\bm{a}\left(\theta,\phi \right) = \dfrac{1}{\sqrt{N_{\rm v}N_{\rm h}}}\bm{a}_{\rm v}\left(\theta\right)\otimes\bm{a}_{\rm h}\left(\theta,\phi\right), \label{eq:2}
\end{equation}
where $\bm{a}_{\rm v}\left(\theta\right)$ and $\bm{a}_{\rm h}\left(\theta,\phi\right)$ are expressed as
\begin{align}
	\bm{a}_{\rm v}\left(\theta\right) &= \left[1,e^{\frac{-j2\pi d_{\rm v}}{\lambda}\cos\theta},\cdots,e^{\frac{-j2\pi\left( N_{\rm v}-1\right)d_{\rm v}}{\lambda}\cos\theta}\right]^{T}
	 , \nonumber \\
	\bm{a}_{\rm h}\left(\theta,\phi\right) &= \left[1,e^{\frac{-j2\pi d_{\rm h}}{\lambda}\sin\theta\cos\phi},\cdots,e^{\frac{-j2\pi\left( N_{\rm h}-1\right)d_{\rm h}}{\lambda}\sin\theta\cos\phi}\right]^{T}\hspace{-0.5em}, \nonumber 
\end{align}
respectively, with $\lambda$ being the carrier wavelength. Then, the time-domain channel shown in \eqref{eq:1} can be transformed into frequency domain by means of the discrete Fourier transform (DFT), that is, 
\begin{align}
		\bm{h}\left[k\right]&=\sum_{l=1}^{L_{\rm d}}\bm{h}\left(l\right)e^{-j\frac{2\pi \left(l-1\right)}{K}k}=\sum_{l=1}^{L_{\rm d}}\bm{A}\bm{g}\left(l\right)e^{-j\frac{2\pi\left(l-1\right)}{K}k} \nonumber \\
		&=\bm{A}\sum_{l=1}^{L_{\rm d}}\bm{g}\left(l\right)e^{-j\frac{2\pi\left(l-1\right)}{K}k}
		=\bm{A}\bm{g}_{k},\ k=1, \cdots, K, \label{eq:3}
\end{align}
where $\bm{h}\left(l\right) = \bm{A}\bm{g}\left(l\right)$ with $\bm{A} \triangleq [\bm{a}(\theta_{1},\phi_{1} ),\cdots,\bm{a}(\theta_{N_{\rm p}},\phi_{N_{\rm p}})]$ composed of $ N_{\rm p} $ array steering vectors, and $\bm{g}\left(l\right)$ is a ${N_{\rm p} \times 1}$ complex-valued vector with $\left[\bm{g}\left(l\right)\right]_{i} = g_{i}f_{\rm rc}\left(lT_{\rm s}-\tau_{i}\right)$, and $\bm{g}_{k}= \sum_{l=1}^{L_{\rm d}}\bm{g}\left(l\right)e^{-j\frac{2\pi\left(l-1\right)}{K}k}$ is the complex-valued channel gain vector pertaining to the $k$-th subcarrier.

It is evident from \eqref{eq:3} that $\bm{g}_{k}$ depends on the subcarrier index $k$ if $L_{\rm d} > 1$ and, thus, the channel is frequency selective. In contrast, the array steering matrix $\bm{A}$ is independent of subcarrier indices, which implies the channels on different subcarriers have the same sparse structure in the beamspace, i.e., identical support. 


%
%
\section{Sparse Channel Estimation}
\label{Section-III}
This section first formulates the problem of sparse channel estimation in the hybrid FD-MIMO system shown in Fig.~\ref{fig:1}. Then, we design a dictionary based on SFG, followed by a Bayesian inference-aided simultaneous orthogonal matching pursuit (BSOMP) algorithm for channel estimation.

\subsection{Problem Formulation}
As the pilots from different users are orthogonal with one another, it is sufficient to consider one single-antenna user when uplink channel estimation is performed. Suppose that $M$ all-one pilots are sent from a user with transmit (Tx) power $P$, the received signal pertaining to the $k$-th subcarrier can be expressed as \cite[Eq. (9)]{SWOMP2018Javier}
\begin{align}
\bm{y}_{\rm w}^{\left(m\right)}\left[k\right]&=\bm{W}_{\rm RF}^{\left(m\right)H}\left(\bm{h}\left[k\right]\sqrt{P} + \bm{n}^{\left(m\right)}\left[k\right]\right) \nonumber \\
&=\sqrt{P}\bm{W}_{\rm RF}^{\left(m\right)H}\bm{h}\left[k\right] + \bm{n}_{\rm w}^{\left(m\right)}\left[k\right], \label{eq:4}
\end{align}
for all $m = 1,\cdots,M$, where $\bm{h}\left[k\right]\in\mathbb{C}^{N_{a}\times 1}$ denotes the channel pertaining to the $k$-th subcarrier, which remains unvaried during the $M$ consecutive pilots; $ \bm{n}^{\left(m\right)}\left[k\right]\sim\mathcal{CN}\left(\bm{0},\sigma^{2}\bm{I}_{N_{a}}\right)$ is the additive white Gaussian noise (AWGN), and $\bm{n}_{\rm w}^{\left(m\right)}\left[k\right] = \bm{W}_{\rm RF}^{\left(m\right)H}\bm{n}^{\left(m\right)}\left[k\right]$ is the combined noise, with $\mathbb{E}\left\lbrace\bm{n}_{\rm w}^{\left(m\right)}\left[k\right]\bm{n}_{\rm w}^{\left(m\right)H}\left[k\right]\right\rbrace = \sigma^{2}\bm{W}_{\rm RF}^{\left(m\right)H}\bm{W}_{\rm RF}^{\left(m\right)}$. Then, the $M$ received pilots in \eqref{eq:4} are concatenated in column as follows:
\begin{equation}
\bm{y}_{\rm w}\left[k\right]=\bm{W}^{H}\bm{h}\left[k\right] + \bm{n}_{w}\left[k\right],\ k=1,\cdots,K, \label{eq:5}
\end{equation}
where
\begin{align}
	\bm{y}_{\rm w}\left[k\right] &= \left[(\bm{y}_{\rm w}^{\left(1\right)}\left[k\right])^H, \cdots, (\bm{y}_{\rm w}^{\left(M\right)}\left[k\right])^H\right]^H \in \mathbb{C}^{ML\times 1}, \label{eq:6}\\ 
	\bm{W} &= \sqrt{P}\left[\bm{W}_{\rm RF}^{\left(1\right)},\cdots,\bm{W}_{\rm RF}^{\left(M\right)}\right]\in\mathbb{C}^{N_{a}\times ML}, \label{eq:7}
\end{align}
and
\begin{equation}
\bm{n}_{\rm w}\left[k\right] = \left[(\bm{n}_{\rm w}^{\left(1\right)}\left[k\right])^H, \cdots, (\bm{n}_{\rm w}^{\left(M\right)}\left[k\right])^H\right]^{H} \in \mathbb{C}^{ML\times 1}.  \label{eq:8}
\end{equation}

It is clear that the covariance matrix of $\bm{n}_{w}\left[k\right]$ given by \eqref{eq:8} is block-diagonal. By recalling the Cholesky factorization, we get the decomposition $\bm{C} = \bm{D}\bm{D}^{H}$, where $\bm{D}$ is a diagonal matrix. Then, by using $\bm{D}^{-1}$ as a whitening filter and inserting \eqref{eq:3} into \eqref{eq:5}, the received pilots can be postprocessed as
\begin{align}
	\bm{y}_{k}
	&= \bm{D}^{-1}\bm{y}_{\rm w}\left[k\right]  \label{eq:9}\\
	&= \bm{D}^{-1}\bm{W}^{H}\bm{A}\bm{g}_{k}+\bm{D}^{-1}\bm{n}_{w}\left[k\right]  \label{eq:10}\\
	&= \bm{\Phi}\bm{A}\bm{g}_{k}+\bm{n}_{k} \label{eq:11}\\
	&\approx \bm{\Phi}\tilde{\bm{A}}\tilde{\bm{g}}_{k}+\bm{n}_{k},\ k=1,\cdots,K, \label{eq:12}
\end{align}
where $\bm{\Phi} \triangleq \bm{D}^{-1}\bm{W}^{H}$ is widely known as sensing matrix, $\bm{n}_{k} \triangleq \bm{D}^{-1}\bm{n}_{w}\left[k\right]$ is the whitened noise. Also, \eqref{eq:12} is obtained by approximating the channel in a sparse angular domain that is defined by an over-complete dictionary $\tilde{\bm{A}} \in \mathbb{C}^{N_{a} \times G}$, where $\tilde{\bm{g}}_{k} \in \mathbb{C}^{G\times1}$ is a sparse channel gain vector to be estimated, with $G$ being the size of the dictionary. Next, we design $\tilde{\bm{A}}$.

\begin{algorithm}[!t]
	\caption{Dictionary Generation based on the SFG} 
	\label{algo:1}
	\begin{algorithmic}[1]
	\REQUIRE  Number of directions $G$, an angle range $\xi \in \left(0, \pi\right]$;
		\FOR {$n=1,\cdots,G$}
			\STATE Compute $x_{n} = 1-{2(n-1)\xi}/{(\left(G-1\right)\pi)}$;
			\STATE Compute ${\alpha}_{n}=\sqrt{1-x_{n}^{2}}$;
			\STATE Compute $y_{n}={\alpha}_{n}\cos\left(n\omega\right)$ and $z_{n}={\alpha}_{n}\sin\left(n\omega\right)$;
			\STATE Transform $[x_{n},y_{n},z_{n}]$ into a spherical coordinate and obtain the zenith angle $\theta_{n}$ and the azimuth angle $\phi_{n}$;
		\ENDFOR
	\ENSURE $\mathbb{S}=\left\lbrace(\theta_{1},\phi_{1}),(\theta_{2},\phi_{2}),\cdots,(\theta_{G},\phi_{G})\right\rbrace \label{eq:set}$.
	\end{algorithmic} 
\end{algorithm}

\subsection{Dictionary Design}
\label{Subsection-III-A}
To accurately estimate channels from \eqref{eq:12}, it is imperative to design a suitable dictionary $\tilde{\bm{A}}=\left[\bm{a}\left(\theta_{1},\phi_{1}\right),\cdots,\bm{a}\left(\theta_{G},\phi_{G} \right)\right]$ that depends on the set of the candidate directions of wave-beams, i.e., $\mathbb{S}\triangleq \{\left(\theta_{1},\phi_{1}\right),\cdots,\left(\theta_{G},\phi_{G}\right)\}$. In particular, to enable consistent accuracy of channel estimation, we design a uniform dictionary based on the SFG~\cite{Fibonacci2009Gonz} that yields nearly uniform directions in 3D beamspace. 

Specifically, given the size $G$ of a dictionary, the set of directions $\mathbb{S}$ based on the SFG can be computed as~\cite{Fibonacci2009Gonz}
\begin{align}
	\mathbb{S} = \left\lbrace\left({\theta}_{n}, {\phi}_{n}\right) \mid 
	\theta_{n} = \arcsin\left(1-\frac{2\left(n-1\right)}{G-1}\right)\right., \nonumber \\
	\phi_{n} = {\rm mod}\left(n\omega,2\pi\right),\ n=1,\cdots,G\bigg\rbrace, \label{eq:13}
\end{align}
where $\arcsin(\cdot)$ is the inverse function of the sine function, ${\rm mod}(a,d)$ denotes the remainder of $a$ divided by $d$, and $\omega \triangleq \left(3-\sqrt{5}\right)\pi$ is the Golden angle. In practice, it is necessary to generate SFG on hemispherical surfaces considering the symmetry of UPAs, or on spheres limited to a certain angles dependent on applications. Accordingly, Algorithm~\ref{algo:1} provides a SFG-based method to generate dictionary within the desired angle range.

For comparison purposes, two typical methods to generate dictionaries, known as the uniform sampling of the physical domain (USPD) and the uniform sampling of the virtual domain (USVD) \cite{offgrid2020Anjinappa}, are reproduced, i.e.,
\begin{align}
	\mathbb{S}_{\rm USPD} = \left\lbrace\left({\theta}_{p},{\phi}_{q}\right)\mid{\theta}_{p}=\frac{\left(p-1\right)\pi}{G_{\rm v}},\ p=1,\cdots,G_{\rm v},\right. \nonumber \\
		\left.{\phi}_{q} = \frac{\left(q-1\right)\pi}{G_{\rm h}},\ q=1,\cdots, G_{\rm h}\right\rbrace, \label{eq:14}
\end{align}
and 
\begin{align}
	\mathbb{S}_{\rm USVD} &= \left\lbrace\left({\theta}_{p},{\phi}_{q}\right)\mid\cos({\theta}_{p})=1-\frac{2p-1}{G_{\rm v}},\ p=1,\cdots,G_{\rm v},\right. \nonumber \\
		&\qquad \left.\cos({\phi}_{q}) = 1-\frac{2q-1}{G_{\rm h}},\ q=1,\cdots, G_{\rm h}\right\rbrace, \label{eq:15}
\end{align}
respectively, where the size of dictionary is $G \triangleq G_{\rm v}G_{\rm h}$.

\begin{figure}[!t]
	\centering
	\vspace{-16pt}
	\subfloat[USPD.]{\includegraphics[width=0.27\linewidth,clip]{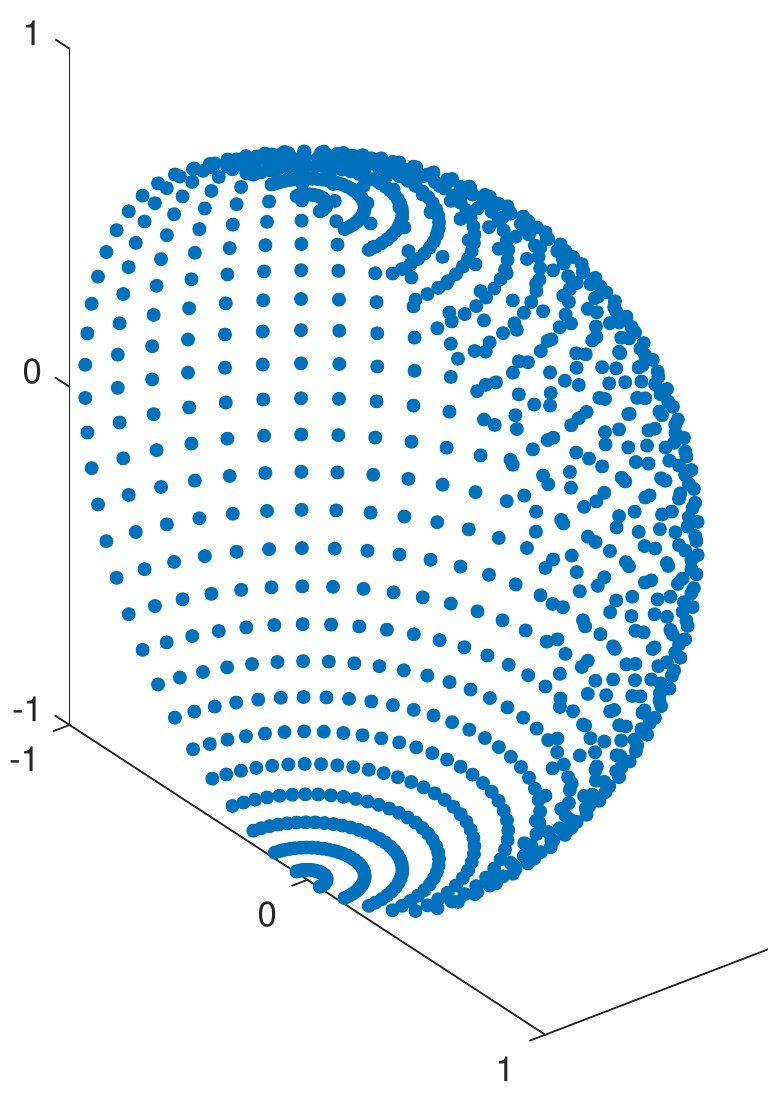}
		\label{fig:2-USPD}}\hfil
	\subfloat[USVD.]{\includegraphics[width=0.27\linewidth,clip]{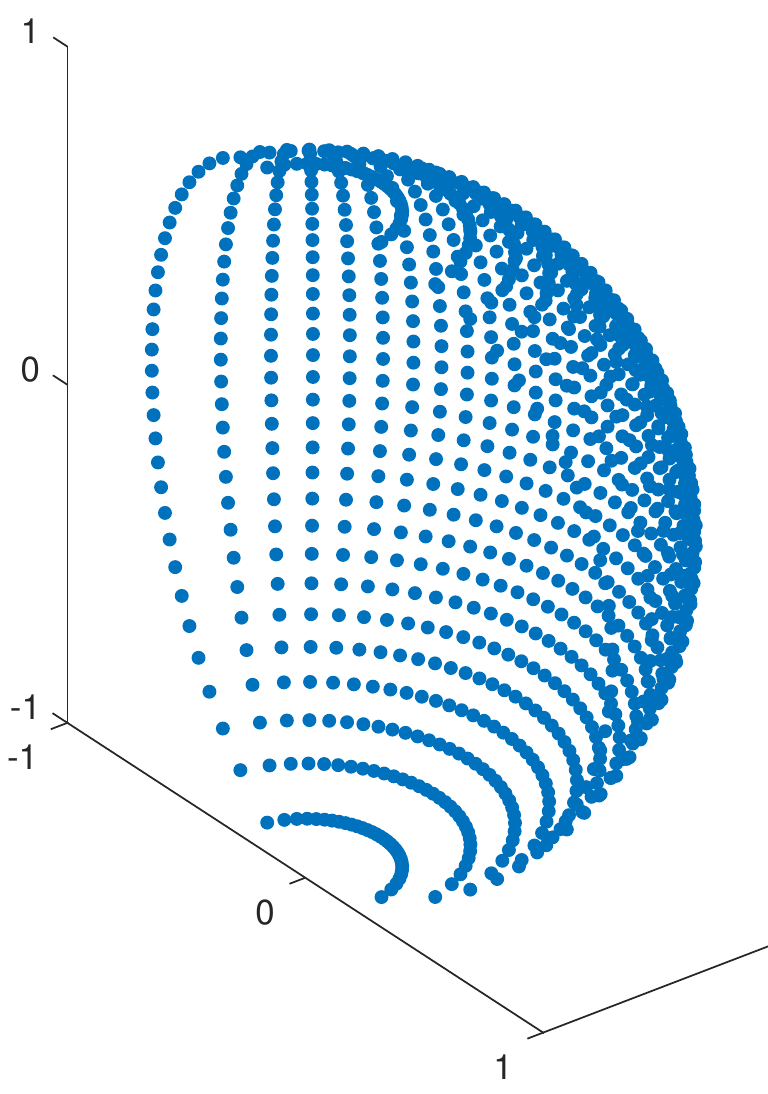}
		\label{fig:2-USVD}}\hfil
	\subfloat[SFG.]{\includegraphics[width=0.27\linewidth,clip]{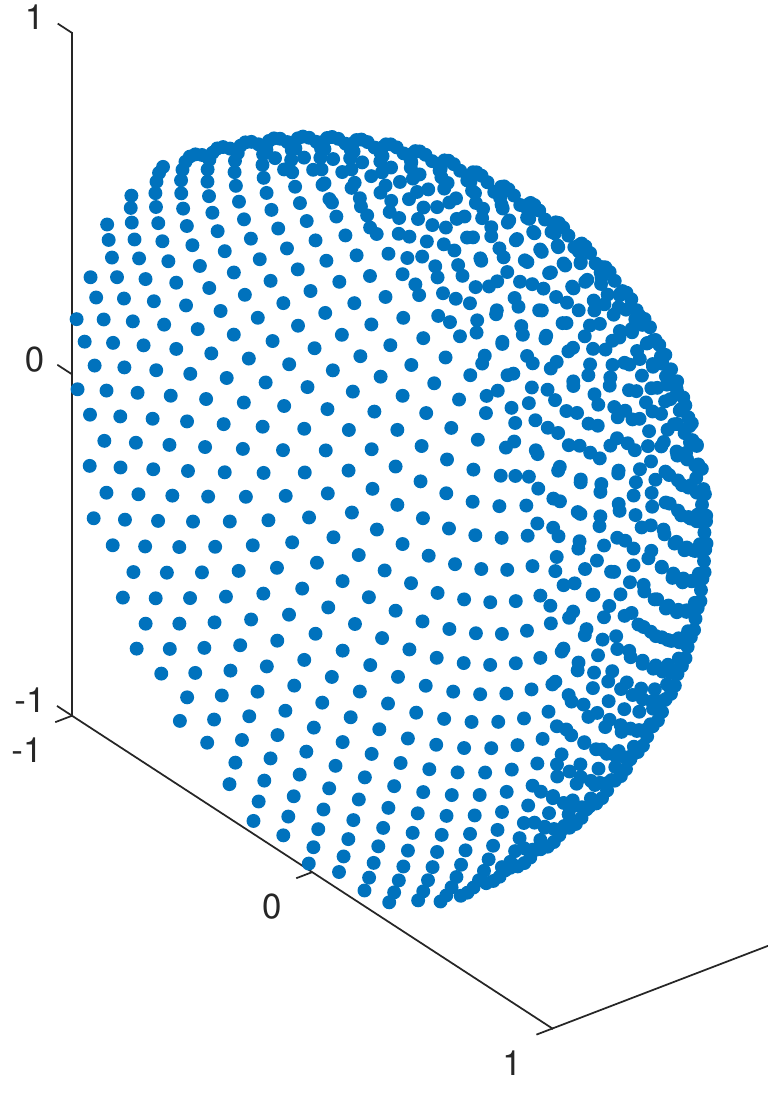}
		\label{fig:2-SFG}}
	\caption{Directional grids on a hemispherical surface, with $G_{\rm v}=G_{\rm h}=32$ and $\xi = \pi/2$.}
	\vspace{-10pt}
	\label{fig:2}
\end{figure}

Figure~\ref{fig:2} depicts the corresponding directional grids generated by the three different methods described above. It is observed from Figs.~\ref{fig:2-USPD} and \ref{fig:2-USVD} that the grids generated by the USPD and USVD are nonuniform whereas, as shown in Fig.~\ref{fig:2-SFG}, the SFG  provides nearly uniform directions in the 3D beamspace. Therefore, applying these dictionaries to estimate channels, the SFG can always obtain consistent estimation accuracy for rays in any direction, unlike the USPD and USVD methods that may fail in the directions with sparse grids \cite{ompce2016Lee}.

\subsection{Bayesian Inference-Aided Channel Estimation}
\label{Subsection-III-B}
After determining the dictionary $\tilde{\bm{A}}$, we are in a position to estimate $\left\lbrace\tilde{\bm{g}}_{k}\right\rbrace_{k=1}^{K}$ in \eqref{eq:12}. Without loss of generality,  we consider the Gaussian prior probability assigned to $\left\lbrace\tilde{\bm{g}}_{k}\right\rbrace_{k=1}^{K}$~as 
\begin{equation}
	P\left(\tilde{\bm{g}}_{k}\right)=\prod_{i=1}^{G}\dfrac{1}{\pi \gamma_{i}}\exp\left(-\frac{\left|\left[\tilde{\bm{g}}_{k}\right]_{i}\right|^{2}}{\gamma_{i}}\right),\ k=1,\cdots,K, \label{eq:16}
\end{equation}
i.e., $\tilde{\bm{g}}_{k}\sim\mathcal{CN}\left(\bm{0},\bm{\Gamma}\right)$, where $\bm{\Gamma}={\rm diag}\left(\gamma_{1},\cdots,\gamma_{G}\right)$. Then, by using the minimum mean squared error (MMSE) criterion, the optimal estimate of $\bm{G}\triangleq \left[ \tilde{\bm{g}}_{1},\cdots,\tilde{\bm{g}}_{K}\right]$ is given by~\cite[Eq.~(11.6)]{booksparse}
\begin{equation}
	\hat{\bm{G}}=\int_{\bm{G}}\bm{G}P\left(\bm{G}|\bm{Y}\right){\rm d}\bm{G}, \label{eq:17}
\end{equation}
where $P\left(\bm{G}|\bm{Y}\right)$ is the posterior probability of $\bm{G}$ given measurements $\bm{Y}\triangleq \left[\bm{y}_{1},\cdots,\bm{y}_{K}\right]$. Since $\left\lbrace\tilde{\bm{g}}_{k}\right\rbrace_{k=1}^{K}$ are sparse and their supports are identical, i.e., $\left\|\tilde{\bm{g}}_{1}\right\|_{0}=\cdots=\left\|\tilde{\bm{g}}_{K}\right\|_{0}=|\mathcal{I}|$ for a support $\mathcal{I}$, $P\left(\bm{G}|\bm{Y}\right)$ can be calculated by using the total probability formula, that is,
\begin{equation}
	P\left(\bm{G}|\bm{Y}\right)=\sum\limits_{\mathcal{I}\in\mathit{\Omega}}P\left(\bm{G}|\mathcal{I},\bm{Y}\right)P\left(\mathcal{I}|\bm{Y}\right), \label{eq:18}
\end{equation}
where $\mathit{\Omega}$ is the set of all possible supports, $P\left(\bm{G}|\mathcal{I},\bm{Y}\right)$ is the posterior probability of $\bm{G}$ given $\mathcal{I}$ and $\bm{Y}$, and $P\left(\mathcal{I}|\bm{Y}\right)$ is the posterior probability of $\mathcal{I}$ given $\bm{Y}$.
Substituting \eqref{eq:18} into \eqref{eq:17} yields
\begin{equation}
	\hat{\bm{G}}=\sum\limits_{\mathcal{I}\in\mathit{\Omega}}P\left(\mathcal{I}|\bm{Y}\right)\int_{\bm{G}}\bm{G}P\left(\bm{G}|\mathcal{I},\bm{Y}\right){\rm d}\bm{G}, \label{eq:19}
\end{equation}
where the inner integral denotes the MMSE estimation of $\bm{G}$ with respect to a known support $\mathcal{I}$ while the outer summation reflects that the final MMSE estimation $\hat{\bm{G}}$ is a weighted average of MMSE estimations over different supports. 

For notational simplicity, we use $\bm{G}_{\mathcal{I}}$ to denote a row-sparse matrix of size $G \times K$ that corresponds to $\bm{G}$ given the support~$\mathcal{I}$. Also, $\bm{Q}_{\mathcal{I}}$ is defined as a sub-matrix of size $|\mathcal{I}|\times K$ that corresponds to the non-zero rows in $\bm{G}_{\mathcal{I}}$. Similarly, $\tilde{\bm{A}}_{\mathcal{I}}$ is a sub-matrix of size $N_{a}\times|\mathcal{I}|$ that corresponds to the support $\mathcal{I}$ in $\tilde{\bm{A}}$ and $\bm{\Gamma}_{\mathcal{I}}$ is a sub-matrix of size $|\mathcal{I}|\times|\mathcal{I}|$ that corresponds to the support $\mathcal{I}$ in $\bm{\Gamma}$. As a result, by recalling the Bayesian formula, the term $P\left(\mathcal{I}|\bm{Y}\right)$ in \eqref{eq:19} can be calculated as
\begin{align}
	\lefteqn{P\left(\mathcal{I}|\bm{Y}\right)=\dfrac{P\left(\bm{Y}|\mathcal{I}\right)P\left(\mathcal{I}\right)}{P\left(\bm{Y}\right)}}  \label{eq:20} \\
		&= \dfrac{P\left(\mathcal{I}\right)}{P\left(\bm{Y}\right)}\int_{\bm{Q}_{\mathcal{I}}}P\left(\bm{Y}|\bm{Q}_{\mathcal{I}},\mathcal{I}\right)P\left(\bm{Q}_{\mathcal{I}}|\mathcal{I}\right)\mathrm{d}\bm{Q}_{\mathcal{I}} \label{eq:21} \\
		&= C_{0}\int_{\bm{Q}_{\mathcal{I}}} \hspace{-1em}C_{1}\exp\hspace{-0.2em}\left(-\frac{1}{\sigma^{2}}\mathrm{tr}\left(\hspace{-0.25em}\left(\bm{Y}-\bm{\Phi}\tilde{\bm{A}}_{\mathcal{I}}\bm{Q}_{\mathcal{I}}\right)^{H}\hspace{-0.5em}\left(\bm{Y}-\bm{\Phi}\tilde{\bm{A}}_{\mathcal{I}}\bm{Q}_{\mathcal{I}}\right)\hspace{-0.25em}\right)\right. \nonumber \\
			&\quad -\mathrm{tr}\left(\bm{Q}_{\mathcal{I}}^{H}\bm{\Gamma}_{\mathcal{I}}^{-1}\bm{Q}_{\mathcal{I}}\right)\hspace{-0.25em}\bigg)\mathrm{d}\bm{Q}_{\mathcal{I}} \label{eq:22} \\
		&= C_{0}C_{1}C_{2}C_{3}\exp\left(\mathrm{tr}\left(\bar{\bm{Q}}_{\mathcal{I}}^{H}\bm{\Sigma}^{-1}\bar{\bm{Q}}_{\mathcal{I}}\right)\right)\det\left(\bm{\Sigma}\right)^{K}, \label{eq:23}
\end{align}
where \eqref{eq:21} is derived by the total probability formula; \eqref{eq:22} is attained by using 
$C_{0} \triangleq P\left(\mathcal{I}\right)/P\left(\bm{Y}\right)$, 
$C_{1} \triangleq \pi^{-|\mathcal{I}|MLK^{2}}\det\left(\bm{\Gamma}_{\mathcal{I}}\right)^{-K}\det\left(\sigma^{2}\bm{I}_{ML}\right)^{-K}$, and the facts that 
$\bm{Y} \sim \mathcal{CN}_{ML,K }\left(\bm{\Phi}\tilde{\bm{A}}_{\mathcal{I}}\bm{Q}_{\mathcal{I}},\bm{I}_{K}\otimes\sigma^{2}\bm{I}_{ML}\right)$ given $\bm{Q}_{\mathcal{I}}$ and $\mathcal{I}$
and 
$\bm{Q}_{\mathcal{I}}\sim\mathcal{CN}_{|\mathcal{I}|,K}\left(\bm{0},\bm{I}_{K}\otimes\bm{\Gamma}_{\mathcal{I}}\right)$ given a support $\mathcal{I}$;\footnote{$\mathcal{CN}_{n,p}\left(\bm{M},\bm{\Upsilon}_{p}\otimes\bm{\Lambda}_{n}\right)$ denotes a Gaussian distribution of matrix variate as defined in \cite[Eq. (2)]{Adhikari2012Matrix}.}
\eqref{eq:23} is derived by rearranging the terms in the exponent and performing integration over $\bm{Q}_{\mathcal{I}}$, with $C_{2} \triangleq \exp\left(-\sigma^{2}\mathrm{tr}\left(\bm{Y}^{H}\bm{Y}\right)\right)$, $C_{3} \triangleq \pi^{|\mathcal{I}|K}$, and $\bm{\Sigma}$ and $\bar{\bm{Q}}_{\mathcal{I}}$ given by, respectively,
\begin{align}
	\bm{\Sigma}& \triangleq \left(\sigma^{-2}\tilde{\bm{A}}_{\mathcal{I}}^{H}\bm{\Phi}^{H}\bm{\Phi}\tilde{\bm{A}}_{\mathcal{I}}+\bm{\Gamma}_{\mathcal{I}}^{-1}\right)^{-1}, \label{eq:24} \\
	\bar{\bm{Q}}_{\mathcal{I}}& \triangleq \sigma^{-2}\bm{\Sigma}\tilde{\bm{A}}_{\mathcal{I}}^{H}\bm{\Phi}^{H}\bm{Y}. \label{eq:25}
\end{align}
For notational simplicity, \eqref{eq:23} can be equivalently expressed~as
\begin{align} \hspace{-0.65em}
	P\left(\mathcal{I}|\bm{Y}\right) = C\exp{\left(\mathrm{tr}\left(\bar{\bm{Q}}_{\mathcal{I}}^{H}\bm{\Sigma}^{-1}\bar{\bm{Q}}_{\mathcal{I}}\right)+K\log\left(\det\left(\bm{\Sigma}\right)\right)\right)}, \hspace{-0.35em} \label{eq:26}
\end{align}
where $C\triangleq C_{0}C_{1}C_{2}C_{3}$ is a constant.
Then, taking the non-zero rows of ${\bm{G}}_{\mathcal{I}}$, i.e., ${\bm{Q}}_{\mathcal{I}}$, into account, the integral term involved in \eqref{eq:19} can be explicitly computed as
\begin{align}
\hat{\bm{Q}}_{\mathcal{I}}&=\int_{\bm{Q}_{\mathcal{I}}}\bm{Q}_{\mathcal{I}}P\left(\bm{Q}_{\mathcal{I}}|\mathcal{I},\bm{Y}\right){\rm d}\bm{Q}_{\mathcal{I}} \label{eq:27} \\
&=\left(\sigma^{-2}\tilde{\bm{A}}_{\mathcal{I}}^{H}\bm{\Phi}^{H}\bm{\Phi}\tilde{\bm{A}}_{\mathcal{I}}+\bm{\Gamma}_{\mathcal{I}}^{-1}\right)^{-1}\sigma^{-2}\tilde{\bm{A}}_{\mathcal{I}}^{H}\bm{\Phi}^{H}\bm{Y}. \label{eq:28}
\end{align}
Hence, the row-sparse matrix $\hat{\bm{G}}_{\mathcal{I}}$ can be recovered by putting $\hat{\bm{Q}}_{\mathcal{I}}$ back as its non-zero rows specified by the support~$\mathcal{I}$. 

In the special case of $|\mathcal{I}|=1$, $P\left(\mathcal{I}|\bm{Y}\right)$ in~\eqref{eq:26} reduces to
\begin{equation}
	 P\left(i|\bm{Y}\right)=C\exp\left(\dfrac{\mathrm{tr}\left(\bm{Y}^{H}\bm{\Psi}_{i}\bm{\Psi}_{i}^{H}\bm{Y}\right)}{\sigma^{4}b}-K\log\left(b\right)\right),  \label{eq:29}
\end{equation}
where $\bm{\Psi}_{i} \triangleq \bm{\Phi}\bm{a}\left(\theta_{i},\phi_{i}\right)$ and $b \triangleq \bm{\Psi}_{i}^{H}\bm{\Psi}_{i}/\sigma^{2}+1/\gamma_{i}$. In the general case of $|\mathcal{I}|>1$, $P\left(\mathcal{I}|\bm{Y}\right)$ can be approximated from \eqref{eq:29} by a greedy pursuit algorithm. As a result, for the MMSE estimate shown in \eqref{eq:19}, we need to compute \eqref{eq:28} and \eqref{eq:29} for all $\mathcal{I}\in\mathit{\Omega}$. In summary, similar to \cite{booksparse}, assuming that $\gamma_{1}=\cdots=\gamma_{G}=\gamma$, i.e., $\bm{\Gamma} \triangleq \gamma\bm{I}$, our Bayesian inference-aided channel estimation method is formalized in Algorithm~\ref{algo:2}. 

\begin{algorithm}[!t]
	\caption{Bayesian Inference-Aided Channel Estimation} 
	\label{algo:2}
	\begin{algorithmic}[1]
	\REQUIRE The received pilots $\bm{Y}$, the sensing matrix $\bm{\Phi}$, the dictionary $\tilde{\bm{A}}$, the noise variance $\sigma^{2}$, and the number of different supports to be considered $V$; 
		\FOR {$v=1,\cdots,V$}
			\STATE {\bf Initialization:} $\mathcal{I}_{v}=\emptyset$ and $\bm{R}_{v}=\bm{Y}$;
			\REPEAT
				\STATE Draw an index $i_{\rm r}\in\left\lbrace1,\cdots,G\right\rbrace\backslash\mathcal{I}_{v} $ randomly with probability proportional to \eqref{eq:29};
				\STATE Update the support by $\mathcal{I}_{v}:=\mathcal{I}_{v}\cup\left\lbrace i_{\rm r}\right\rbrace$;
				\STATE Update the selected dictionary $\tilde{\bm{A}}_{\mathcal{I}_{v}}$;
				\STATE Obtain $\hat{\bm{Q}}_{\mathcal{I}_{v}}=\left(\tilde{\bm{A}}_{\mathcal{I}_{v}}^{H}\tilde{\bm{A}}_{\mathcal{I}_{v}}\right)^{-1}\tilde{\bm{A}}_{\mathcal{I}_{v}}^{H}\bm{Y}$;
				\STATE Update the residual matrix by $\bm{R}_{v}=\bm{Y}-\tilde{\bm{A}}_{\mathcal{I}_{v}}\hat{\bm{Q}}_{\mathcal{I}_{v}}$;
			\UNTIL {$\frac{1}{MLK}\left\|\bm{R}_{v}\right\|_{\rm F}^{2}\leq\sigma^{2}$};
			\STATE Compute $\bm{\Gamma}_{\mathcal{I}_{v}} = \frac{1}{|\mathcal{I}_{v}|K}\left\|\hat{\bm{Q}}_{\mathcal{I}_{v}}\right\|_{\rm F}^{2}\bm{I}$;
			\STATE Refine the estimation of $\hat{\bm{Q}}_{\mathcal{I}_{v}}$ by \eqref{eq:28};  
			\STATE Recover $\hat{\bm{G}}_{\mathcal{I}_{v}}$ from $\hat{\bm{Q}}_{\mathcal{I}_{v}}$ by treating the rows of the latter as the non-zero rows of the former;
		\ENDFOR
		\STATE Compute $\hat{\bm{G}}=\frac{1}{V}\sum_{v=1}^{V}\hat{\bm{G}}_{\mathcal{I}_{v}}$;
	\ENSURE The estimated channel $\hat{\bm{H}} =\tilde{\bm{A}}\hat{\bm{G}}$.
	\end{algorithmic} 
\end{algorithm}%

\section{Simulation Results and Discussions}
\label{Section-IV}
This section presents and discusses simulation results of channel estimations in terms of the designed dictionary and the support recovery algorithm, by comparing our proposed method with the benchmark ones reported in \cite{ompce2016Lee,SWOMP2018Javier}.

\subsection{The Effectiveness of Dictionary}

To evaluate the effectiveness of dictionary, we explore the statistics of the minimal angle of the angles between a random point on the unit hemisphere and all the $G$ grids in a dictionary. Mathematically, we consider an ideal case and suppose that $G$ grids are uniform distributed on a unit hemisphere and the minimal angle of the angles between a random point on the unit hemisphere and the $G$ grids is no bigger than $r_{0}$. In such a case, the hemispherical surface can be approximately tessellated into $G$ spherical circles with radius $r_{0}$. Accordingly, we have ${2\pi}/{G}=2\pi\left(1-\cos r_{0}\right)$, yielding
\begin{equation}
	r_{0}=\arccos\left(\dfrac{G-1}{G}\right), \label{eq:30}
\end{equation}
where $\arccos(\cdot)$ is the inverse function of the cosine function. Moreover, the cumulative distribution function (CDF) of the theoretically minimal angle can be derived and given by
\begin{equation}
	F\left(r\right)=\dfrac{r^{2}}{r^{2}_{0}},\ 0\leq r\leq r_{0}. \label{eq:31}
\end{equation}

\begin{figure*}[!t]
	\centering
	\begin{minipage}[!t]{0.3\textwidth}
		\centering
		\includegraphics[width=\linewidth]{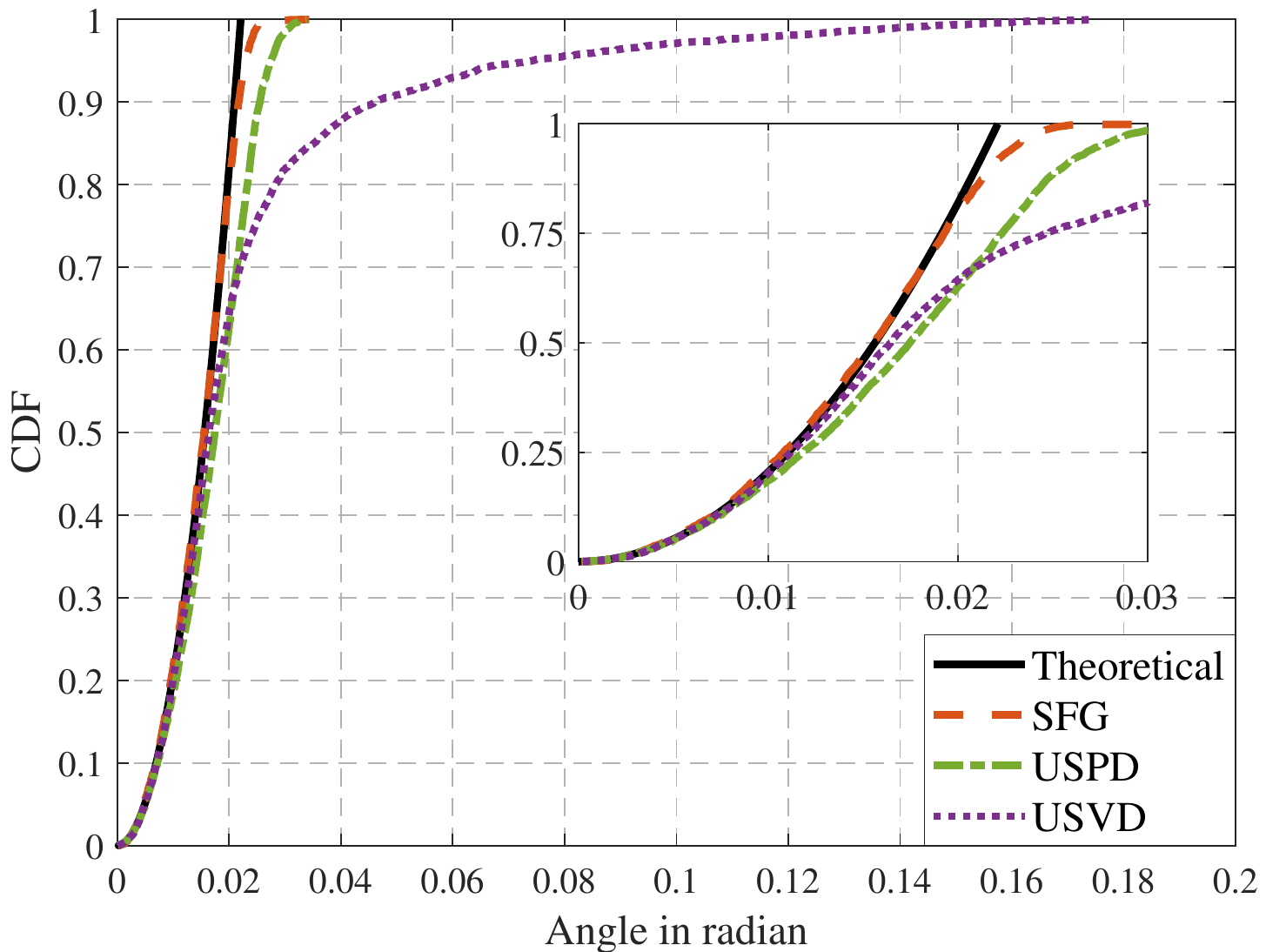}
		\caption{CDFs of the minimal angles of the grids generated by \eqref{eq:13}-\eqref{eq:15}, compared with the theoretical one by \eqref{eq:31}.}
		\label{fig:3}
	\end{minipage}\hfil
	\begin{minipage}[!t]{0.3\textwidth}
		\centering
		\vspace{-9pt}
		\includegraphics[width=\linewidth]{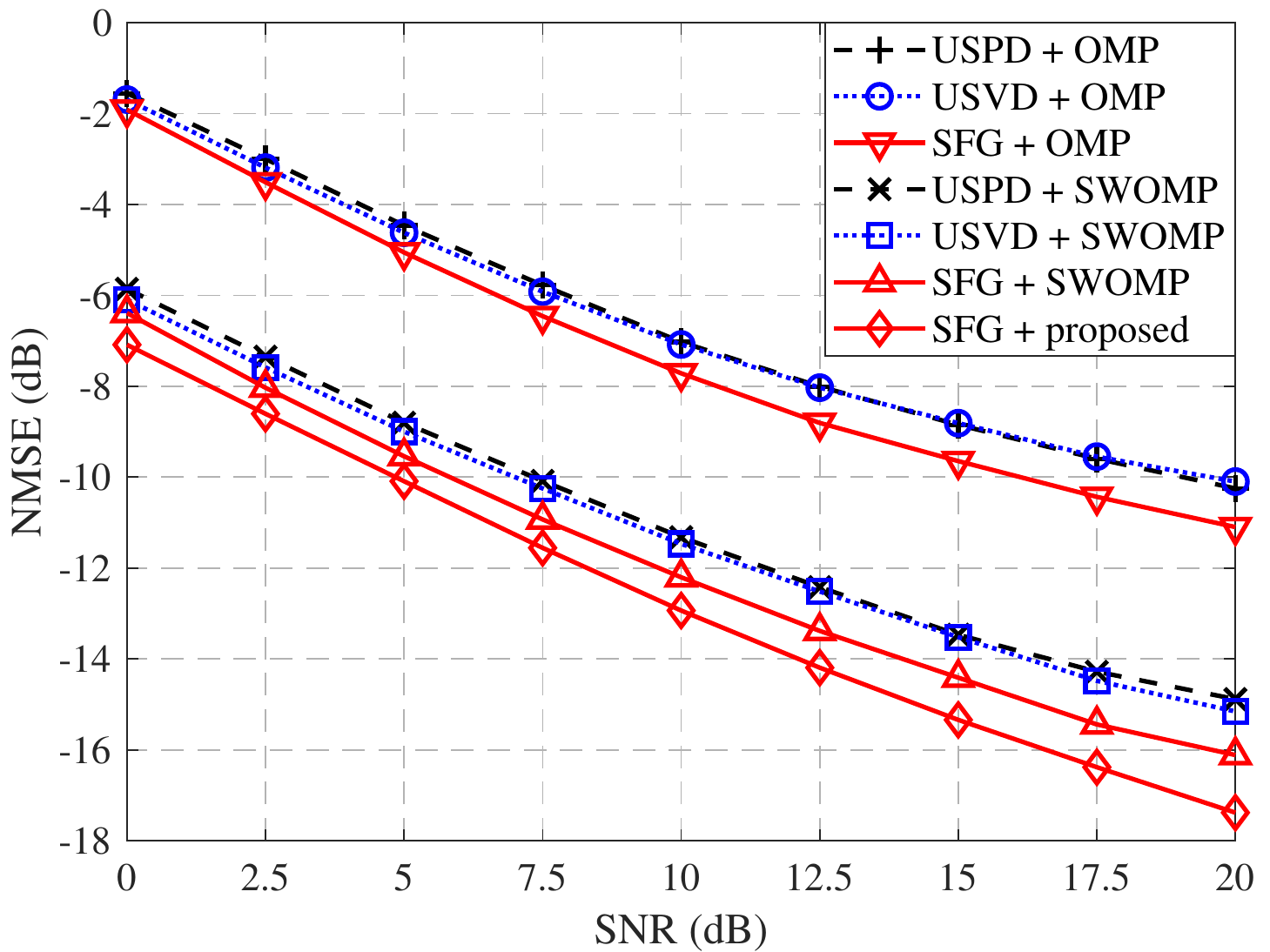}
		\caption{NMSE versus SNR in the case of~$M=10$ pilots.}
		\label{fig:4}
	\end{minipage}\hfil
	\begin{minipage}[!t]{0.3\textwidth}
		\centering
		\vspace{-9pt}
		\includegraphics[width=\linewidth]{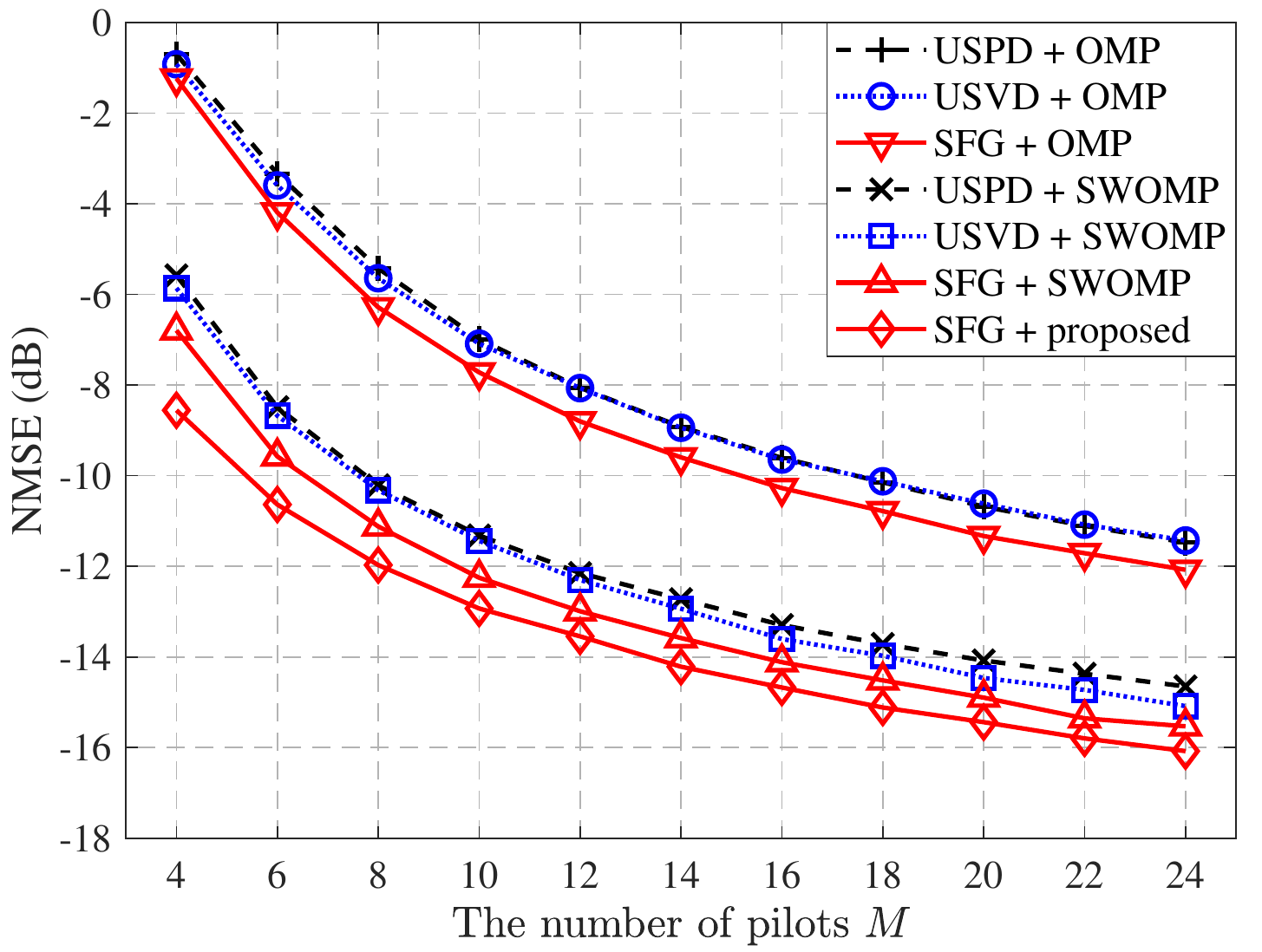}
		\caption{NMSE versus the number of pilots in the case of SNR~$=10$~dB.}
		\label{fig:5}
	\end{minipage}
\end{figure*}

Figure~\ref{fig:3} depicts the numerical results computed by \eqref{eq:31}, compared with the simulation results of the CDF of the minimal angles corresponding to the grids generated by \eqref{eq:13}-\eqref{eq:15}, where $G_{\rm h}=G_{\rm v}=64$. It is observed that the CDF pertaining to the SFG method is closer to the theoretical one than either the UPSD or the USVD method. This observation accords with that from Fig.~\ref{fig:2}, where the grids generated by the proposed SFG method are more uniform than those generated by the UPSD or  USVD. Therefore, the dictionary generated by the SFG method is capable of more consistent channel estimation for rays in any direction.

\subsection{Performance of Channel Estimation}
To evaluate the performance of Algorithm~\ref{algo:2} for channel estimation, in the pertaining simulation experiments we assume that the UPA at the BS is equipped with $ L = 8$ RF chains, $(N_{\rm v}, N_{\rm h}) = (16, 16)$ antenna elements with spacings $d_{\rm v}=d_{\rm h}=\lambda/2$, and $K=24$ subcarriers. The number of spatial scattering paths $N_{\rm p} \in \{6, 7, \cdots, 12\}$ is chosen equiprobably, the directions of paths are isotropic, the gains of each path are subject to zero-mean complex Gaussian distribution such that $\mathbb{E}\left\lbrace\left\|\bm{h}\left[k\right]\right\|_{\mathrm{F}}^{2}\right\rbrace=N_{\rm v}N_{\rm h}$, and the delays $\{\tau_{n}\}$, $n=1, \cdots, N_{\rm p}$, are chosen uniformly from $[0,(N_{\rm p}-1)T_{\rm s}]$ with $T_{\rm s}=1/1760$~$\mu$s~\cite{SWOMP2018Javier}. As for the dictionaries, we set $G_{\rm h}=G_{\rm v}=32$ and $V=20$ is chosen in Algorithm~\ref{algo:2}. As a performance measure, the normalized mean squared error (NMSE) is defined as $\mathbb{E}\left\lbrace\|\hat{\bm{H}}-\bm{H}\|_{\mathrm{F}}^{2}/\left\|\bm{H}\right\|_{\mathrm{F}}^{2}\right\rbrace$, where $\bm{H}$ denotes the actual channel used in simulation experiments. 

Figure~\ref{fig:4} compares the NMSE of different schemes versus the signal-to-noise ratio (SNR), in the case of $M=10$ pilots. For comparison purposes, the three dictionaries generated by \eqref{eq:13}-\eqref{eq:15} are combined each with the traditional OMP or SWOMP scheme, widely used for sparse channel estimation in mmWave MIMO systems. We can observe that, if the OMP scheme is concerned, the first three curves in Fig.~\ref{fig:4} show the `USPD+OMP' scheme performs the worst and the `SFG+OMP' the best, while the `UVSD+OMP' performs in-between. This observation is in agreement with that from Fig.~\ref{fig:3}. The same conclusion can be taken from Fig.~\ref{fig:4} if the SWOMP scheme is concerned. Finally, the last curve in Fig.~\ref{fig:4} illustrates that the proposed Bayesian approach combined with the SFG-based dictionary outperforms the other schemes under consideration. Likewise, we can draw a similar conclusion from Fig.~\ref{fig:5}, where the NMSE is plotted versus the number of pilots with a fixed SNR. Therefore, the proposed SFG-based dictionary and the companion Bayesian approach benefit more accurate channel estimation.

\section{Conclusions}
\label{Section-V}
This letter designed a spherical Fibonacci grid-based dictionary with minor angular errors to accurately represent channels in the 3D beamspace, and developed a Bayesian inference-aided simultaneous orthogonal matching pursuit algorithm for efficient channel estimation in hybrid FD-MIMO communication systems. Numerical results demonstrated that the designed dictionary generates more uniform directional grids in any desired 3D beamspace. The Bayesian channel estimation combined with the designed dictionary yields higher accuracy than the benchmark schemes, which finds applications in mmWave FD-MIMO systems.

\IEEEpubidadjcol

\bibliographystyle{IEEEtran}
\bibliography{IEEEabrv, References}

\begin{thebibliography}{10}
\providecommand{\url}[1]{#1}
\csname url@samestyle\endcsname
\providecommand{\newblock}{\relax}
\providecommand{\bibinfo}[2]{#2}
\providecommand{\BIBentrySTDinterwordspacing}{\spaceskip=0pt\relax}
\providecommand{\BIBentryALTinterwordstretchfactor}{4}
\providecommand{\BIBentryALTinterwordspacing}{\spaceskip=\fontdimen2\font plus
\BIBentryALTinterwordstretchfactor\fontdimen3\font minus
  \fontdimen4\font\relax}
\providecommand{\BIBforeignlanguage}[2]{{%
\expandafter\ifx\csname l@#1\endcsname\relax
\typeout{** WARNING: IEEEtran.bst: No hyphenation pattern has been}%
\typeout{** loaded for the language `#1'. Using the pattern for}%
\typeout{** the default language instead.}%
\else
\language=\csname l@#1\endcsname
\fi
#2}}
\providecommand{\BIBdecl}{\relax}
\BIBdecl

\bibitem{Overview2016Heath}
R.~W. {Heath}, N.~{González-Prelcic}, S.~{Rangan}, W.~{Roh}, and A.~M.
  {Sayeed}, ``An overview of signal processing techniques for millimeter wave
  {MIMO} systems,'' \emph{IEEE J. Sel. Top. Signal Process.}, vol.~10, no.~3,
  pp. 436--453, Apr. 2016.

\bibitem{swomp2018jp}
J.~P. {González-Coma}, J.~{Rodríguez-Fernández}, N.~{González-Prelcic},
  L.~{Castedo}, and R.~W. {Heath}, ``Channel estimation and hybrid precoding
  for frequency selective multiuser mmwave {MIMO} systems,'' \emph{IEEE J. Sel.
  Top. Signal Process.}, vol.~12, no.~2, pp. 353--367, May 2018.

\bibitem{9298895}
H.~Xie, P.~Wu, F.~Tan, and M.~Xia, ``Adaptively-regularized compressive sensing
  with sparsity bound learning,'' \emph{IEEE Commun. Lett.}, vol.~25, no.~4,
  pp. 1283--1287, Apr. 2021.

\bibitem{ompce2016Lee}
J.~{Lee}, G.~{Gil}, and Y.~H. {Lee}, ``Channel estimation via orthogonal
  matching pursuit for hybrid {MIMO} systems in millimeter wave
  communications,'' \emph{IEEE Trans. Commun.}, vol.~64, no.~6, pp. 2370--2386,
  Jun. 2016.

\bibitem{SWOMP2018Javier}
J.~{Rodríguez-Fernández}, N.~{González-Prelcic}, K.~{Venugopal}, and R.~W.
  {Heath}, ``Frequency-domain compressive channel estimation for
  frequency-selective hybrid millimeter wave {MIMO} systems,'' \emph{IEEE
  Trans. Wirel. Commun.}, vol.~17, no.~5, pp. 2946--2960, May 2018.

\bibitem{2017BL}
A.~Mishra, A.~Rajoriya, A.~K. Jagannatham, and G.~Ascheid, ``Sparse {Bayesian}
  learning-based channel estimation in millimeter wave hybrid {MIMO} systems,''
  in \emph{Proc. IEEE SPAWC}, 2017, pp. 1--5.

\bibitem{2018BL}
M.~Umar~Aminu, M.~Codreanu, and M.~Juntti, ``Bayesian learning based
  millimeter-wave sparse channel estimation with hybrid antenna array,'' in
  \emph{Proc. IEEE SPAWC}, 2018, pp. 1--5.

\bibitem{2021BL}
S.~Srivastava, R.~K. Singh, A.~K. Jagannatham, and L.~Hanzo, ``Bayesian
  learning aided simultaneous row and group sparse channel estimation in
  orthogonal time frequency space modulated {MIMO} systems,'' \emph{IEEE Trans.
  Commun. (Early Access)}, 2021.

\bibitem{Fibonacci2009Gonz}
{\'A}.~{Gonz{\'a}lez}, ``{Measurement of areas on a sphere using Fibonacci and
  latitude-longitude lattices},'' \emph{Math. Geosci.}, vol.~42, pp. 49--64,
  Nov. 2009.

\bibitem{booksparse}
M.~{Elad}, \emph{Sparse and Redundant Representations: {From} Theory to
  Applications in Signal and Image Processing}.\hskip 1em plus 0.5em minus
  0.4em\relax New York, NY, USA: Springer, 2010.

\bibitem{offgrid2020Anjinappa}
C.~K. {Anjinappa}, A.~C. {Gürbüz}, Y.~{Yapıcı}, and {\.I}.~{Güvenç},
  ``Off-grid aware channel and covariance estimation in mmwave networks,''
  \emph{IEEE Trans. Commun.}, vol.~68, no.~6, pp. 3908--3921, Jun. 2020.

\bibitem{Phase2016Rial}
R.~{Méndez-Rial}, C.~{Rusu}, N.~{González-Prelcic}, A.~{Alkhateeb}, and R.~W.
  {Heath}, ``Hybrid {MIMO} architectures for millimeter wave communications:
  Phase shifters or switches?'' \emph{IEEE Access}, vol.~4, pp. 247--267, Mar.
  2016.

\bibitem{GBSM2018}
Z.~Lian, L.~Jiang, and C.~He, ``{A 3-D GBSM} based on isotropic and
  non-isotropic scattering for {HAP-MIMO} channel,'' \emph{IEEE Commun. Lett.},
  vol.~22, no.~5, pp. 1090--1093, May 2018.

\bibitem{Adhikari2012Matrix}
Adhikari and Sondipon, ``Matrix variate distributions for probabilistic
  structural dynamics,'' \emph{AIAA Journal}, vol.~45, no.~7, pp. 1748--1762,
  2012.

\end{thebibliography}
\vfill
\end{document}